\input{aipcheck}

\documentclass[
    ,final            
  ]
  {aipproc}

\layoutstyle{6x9}
\newcommand{\lsim}{\stackrel{<}{_\sim}}
\newcommand{\gsim}{\stackrel{>}{_\sim}}
\newcommand{\tppp}{\tau\to\pi\pi\pi\nu_\tau}

\begin{document}

\title{Basics of Resonance Chiral Theory}


\author{J.~Portol\'es}{
  address={Instituto de F\'{\i}sica Corpuscular, IFIC, CSIC-Universitat de Val\`encia, Edifici d'Instituts de Paterna, Apt. Correus 22085, E-46071 Val\`encia, Spain}
}

\begin{abstract}
We review the main components that have to be considered, within Resonance Chiral Theory, in the study of processes whose dynamics is dominated by hadron resonances. We show its application in the study
of the $\tau \rightarrow \pi \pi \pi \nu_{\tau}$ decay.
\end{abstract}

\maketitle


\section{Introduction}
The dynamics prompted by strong interactions in the hadronic low-energy region is driven by
Quantum Chromodynamics (QCD) in its non-perturbative regime. The study of processes involving
the lower part of the hadronic spectrum (namely $E \lsim 2 \, \mbox{GeV}$) is not feasible with
a strong interaction theory written in terms of dynamical quarks and gluons. Theoretically the 
way out would be to trade partonic QCD for a dual theory written in terms of the relevant 
degrees of freedom, i.e. mesons and baryons. However we still do not know such a theory in the full
energy regime pointed out. Only for $E \ll M_{\rho}$ (being $M_{\rho}$ the mass of the $\rho(770)$)
we have an effective field theory of QCD, namely Chiral Perturbation Theory ($\chi$PT)
\cite{Gasser:1983yg,Gasser:1984gg,Leutwyler:1993iq}. 
\par
The study of the phenomenology of hadron processes for $M_{\rho} \lsim E \lsim 2 \, \mbox{GeV}$, 
is carried out, at present, using several approaches. A promising one is, undoubtedly, lattice
gauge calculations. However, on one side it relies in approximations that are far from being under control,
and on the other, observables involving Green functions of four points or more are still not compelled with
the present technology. Hence we have to count on other approaches like ad-hoc parameterizations, 
phenomenological Lagrangian models or, tentatively, effective field theories (EFT). However the construction
of an EFT driven by QCD in the energy region populated by resonances is still an
unsolved problem. 
\par
Essentially the difficulty lies in a key element for the construction of an appropriate EFT~: 
 our lack of knowledge on how to include the
information of higher energies into the couplings of a Lagrangian written in terms of 
the fields that are active in that energy regime, resonances and Goldstone bosons. While 
$\chi$PT arises from the chiral symmetry of massless QCD and the integration of the spectrum
above $M_{\rho}$, the construction of an EFT for $M_{\rho} \lsim E \lsim 2 \, \mbox{GeV}$ lacks
a clear model-independent scenery.
\par
We do not really need a Lagrangian theory to study hadronic processes. The knowledge and construction
of the relevant Green functions is enough for that task. However we think that a Lagrangian is a powerful
tool because it includes, naturally, the relevant symmetries into the game without having to care about
including that information in later stages.
\par
Within the phenomenological Lagrangian approach, the Resonance Chiral Theory (R$\chi$T) setting 
\cite{Ecker:1988te,Ecker:1989yg,Cirigliano:2006hb} incorporates Goldstone bosons
and resonance fields into a Lagrangian frame driven by chiral and unitary symmetry, respectively
\cite{Coleman:1969sm,Callan:1969sn}. The asymptotic behaviour of the Green functions of QCD currents is used both to constrain the number of operators and their couplings. Finally, in order to ease the procedure a model of single resonance approximation within Large-$N_C$ is considered \cite{Ecker:1988te,Peris:1998nj,Golterman:1999au}. 
At present R$\chi$T only involves meson fields. We will recall the method with more detail in the next section where we unravel the state of the art of R$\chi$T and foresee further developments and needs. 
\par
We end this note with an example of application of the methodology in the study of form factors
in $\tau \rightarrow \pi \pi \pi \nu_{\tau}$ decays.

\section{Resonance Chiral Theory}
The lack of a mass gap between the spectrum of light-flavoured meson resonances and the perturbative
continuum (let us say $E \gsim 2 \, \mbox{GeV}$) hinders the construction of an appropriate EFT to 
handle the strong interactions in the energy region populated by those resonances. However there are 
several tools that allow us to grasp relevant features of QCD and to implement them into an
EFT-like Lagrangian model. Two key premises are the following~:
\par
1) A result put forward by S.~Weinberg \cite{Weinberg:1978kz} and worked out by H.~Leutwyler
\cite{Leutwyler:1993iq} states that, if one writes down the most general possible Lagrangian, including
all terms consistent with assumed symmetry principles, and then calculates matrix elements with this 
Lagrangian to a given order in perturbation theory, the result will be the most general possible
S-matrix amplitude consistent with analyticity, perturbative unitarity, cluster decomposition and the
principles of symmetry that have been specified.
\par
2) It has been pointed out \cite{'tHooft:1973jz,Witten:1979kh} that the inverse of the number of 
colours of the $SU(N_C)$ gauge group could be taken as a perturbative expansion parameter. Indeed
Large-$N_C$ QCD shows features that resemble, both qualitative and quantitatively, the $N_C=3$ case.
A relevant consequence of this approach is, for instance, that meson dynamics is described in the
$N_C \rightarrow \infty$ limit by the tree diagrams of an effective local Lagrangian whose spectrum
includes and infinite number of zero-width states.
\par
Both assertions can be merged by constructing a Lagrangian theory in terms of $SU(3)$ (Goldstone
bosons) and $U(3)$ (heavier resonances) flavour multiplets that we call
Resonance Chiral Theory. This has
systematically been established \cite{Ecker:1988te,Ecker:1989yg,Cirigliano:2006hb} and sets 
forth the following features~:
\par
i) We have to consider which spectrum of states do we include in our theory, other than the
lightest octet of pseudoscalar mesons (Goldstone bosons). As we bear with Large-$N_C$ settings we 
should consider an infinite number of states. However this is not feasible in a model-independent way
and we decide to cut the spectrum relying in the lightest multiplets that, it is well known, dominate
the dynamics. Keeping still contact with the leading order in the $1/N_C$ expansion we should include
the lightest multiplets of scalar, pseudoscalar, vector and axial-vector mesons, at least, that remain
in the $N_C \rightarrow \infty$ limit. Though there is a heated debate surrounding the $J=0$ resonant
states, the situation with the $J=1$ mesons, mainly from lattice determinations 
\cite{Grignani:2003uv,Petry:2008rt,Engel:2010my}, seems rather 
settled, leaving the ones including $\rho(770)$ and $\mathrm{a}_1(1260)$ as those vector and axial-vector
nonets that stay in the Large-$N_C$ limit.
\par
ii) The construction of the operators is guided by chiral symmetry for the lightest pseudoscalar
mesons (Goldstone bosons) and by unitary symmetry for the resonances. Typically~:
\begin{equation} \label{eq:rchto}
 {\cal O} \, \sim \, \langle R_1 R_2 ... \chi(p^n) \rangle \, , 
\end{equation}
where $R_i$ indicates a resonance field and $\chi(p^n)$ is a chiral structured tensor, involving
the Goldstone bosons and external currents, with a chiral counting represented by the power of
momenta. With this structure chiral symmetry is preserved upon integration of the resonance states,
and the low-energy expansion of the amplitudes show the appropriate behaviour.
\par
iii) As in $\chi$PT, symmetries do not provide information on the coupling constants that weight
the different operators. These couplings only incorporate information from higher energies. However
if we intend that our theory plays the role of a mediator between the chiral and the parton regimes,
it is clear that the amplitudes or form factors arising from our Lagrangian have to match the 
asymptotic behaviour driven by QCD. A heuristic strategy, well supported by the phenomenology
\cite{Peris:1998nj,Knecht:2001xc,Pich:2002xy,RuizFemenia:2003hm,Cirigliano:2004ue,Cirigliano:2005xn,Portoles:2003ww,Bijnens:2003rc},
lies in matching the Operator Product Expansion (OPE) of Green functions (that are order parameters
 of chiral symmetry) with the
corresponding expressions evaluated within the theory. This procedure provides a determination
for some of the coupling constants of the Lagrangian. In addition, the asymptotic behaviour of form
factors of QCD currents is estimated from the spectral structure of two-point functions or the 
partonic make up \cite{Brodsky:1973kr,Lepage:1980fj}.
\par
iv) The theory that we are devising lacks an
expansion parameter. There is the guide given by Large-$N_C$ that translates into a loop expansion.
However there is no counting that limits the number of operators with resonances that have to be
included in the initial Lagrangian. This non-perturbative character of R$\chi$T, that may take back
the perturbative practitioner, has to be understood properly. On one side the number of resonance
fields relies fundamentally on the physical system of interest, on the other, the maximum order
of the chiral tensor $\chi(p^n)$ in Eq.~(\ref{eq:rchto}) is very much constrained by the enforced
high-energy behaviour, as explained in iii) before. Customarily higher powers of momenta lean to 
spoil the asymptotic bearing that QCD demands.
\par
The processes that take place in the energy region populated by many resonances, $M_{\rho} \lsim E \lsim 2 \, \mbox{GeV}$ are not the only ones that R$\chi$T addresses. In fact at its origins lies the
determination of the low-energy constants (LECs) from $\chi$PT. As
an EFT, the LECs that arise at ${\cal O}(p^n)$ carry the short-distance information of higher
energies mainly driven by the spectrum immediately above the Goldstone bosons, i.e. lightest resonances.
Hence the information of these is carried out into the LECs upon integration of the resonance fields.
In Refs.~\cite{Ecker:1988te,Ecker:1989yg} the R$\chi$T Lagrangian giving the ${\cal O}(p^4)$ $\chi$PT 
Lagrangian was considered and it was shown that, at this order, the chiral LECs are saturated by the resonance
contributions. Later on the same exercise has been carried out at ${\cal O}(p^6)$
\cite{Cirigliano:2006hb}. This task is very important because the state of the art in $\chi$PT involves
already many ${\cal O}(p^6)$ calculations which predictability depends on our knowledge of the
LECs participating in those observables. A lot of effort is employed in their determination
\cite{Kampf:2006bn,Rosell:2007mp,GonzalezAlonso:2008rf,Pich:2008xj,GonzalezAlonso:2009ns}.

\subsection{R$\chi$T~: Application}

The application of R$\chi$T in the study of a particular observable follows a perturbative scheme
driven by Large-$N_C$, i.e. one starts at leading order with the classical Lagrangian and tree-level
diagrams, then one can proceed to one-loop diagrams that are next-to-leading in the $1/N_C$ expansion,
and so on. In this section we will deal with the $N_C \rightarrow \infty$ limit and leave for the 
next section higher-order developments.
\par
Let us consider an specific observable such as a form factor parameterizing the hadronization 
of an external current in an energy region populated by resonances. In general such a form factor 
would be provided by a Green function
of $n$ points ($n \geq 2$). Then we proceed as follows~:
\par
1) The Lagrangian of R$\chi$T is given by~:
\begin{equation} \label{eq:rchtlag}
 {\cal L}_{\mbox{\footnotesize R$\chi$T}} \, = \, {\cal L}_{\mbox{\footnotesize kinR}} \, + 
\, {\cal L}_{\mbox{\footnotesize $\chi$PT}}^R(\phi, {\cal J}) \, + \, {\cal L}_I(\phi,R,{\cal J}) \, ,
\end{equation}
where ${\cal L}_{\mbox{\footnotesize kinR}}$ stands for the kinetic term involving resonances,
$\phi$ stands for the lightest pseudoscalar mesons (Goldstone bosons), $R$ for
the resonances and ${\cal J}$ for external currents, that allow us to evaluate the Green functions in an easy
manner. ${\cal L}_{\mbox{\footnotesize $\chi$PT}}^R$ has the same structure of operators than the 
$\chi$PT Lagrangian ${\cal L}_{\mbox{\footnotesize $\chi$PT}}$ but its LECs are not the same. If, for instance,
one uses the antisymmetric formulation of spin-1 resonances then
${\cal L}_{\mbox{\footnotesize $\chi$PT}}\left( {\cal O}(p^4) \right)$
 has to be excluded in Eq.~(\ref{eq:rchtlag}), i.e. their LECs vanish \cite{Ecker:1989yg}.
It is by integrating out the resonance states in ${\cal L}_{\mbox{\footnotesize R$\chi$T}}$ that we
obtain ${\cal L}_{\mbox{\footnotesize $\chi$PT}}$.
\par
2) We construct the interaction Lagrangian with all the {\em simplest} operators ${\cal O}$ in Eq.~(\ref{eq:rchto}) 
that are given by chiral symmetry (for the Goldstone bosons) and unitary symmetry (resonance fields), 
contributing to the relevant Green functions \cite{Ecker:1988te,Ecker:1989yg,Cirigliano:2006hb,Kampf:2006yf}~:
\begin{equation} \label{eq:li}
 {\cal L}_I \, = \, \sum_{i} \, \alpha_i \, {\cal O}^i\left( \phi, R , {\cal J} \right) \, .
\end{equation}
Here the set $\{\alpha_i\}$ indicates unknown couplings that are not fixed by symmetry requirements
alone. By the {\em simplest} operators we mean those with the 
minimal number of derivatives on whatever fields.
\par
3) Once the Lagrangian theory is constructed we can give an analytical expression for the form factor in 
terms of the $\{ \alpha_i \}$ unknown couplings. This expression is the most general
parameterization satisfying chiral and unitary symmetries and analyticity. At this point though, we are not
predictive because our ignorance on the couplings. Indeed we could use our result to fit the couplings from 
data and terminate the procedure here. However this would transfer poor information from QCD into our theory.
We can do better as we know explain.
\par
4) We consider Green functions of QCD currents that involve the couplings in our Lagrangian. We evaluate
the OPE of the Green functions within partonic QCD at leading order and, analogously, we perform the same 
evaluation with our Lagrangian theory. If the Green function is an order parameter of 
chiral symmetry (i.e. has no one-loop partonic contributions in the chiral limit) the matching between
the OPE expansion and our evaluation within R$\chi$T is directly feasible 
\cite{Peris:1998nj,Knecht:2001xc,Pich:2002xy,RuizFemenia:2003hm,Cirigliano:2004ue,Cirigliano:2005xn,Portoles:2003ww,Bijnens:2003rc}.
If they are not order parameters the situation is more involved \cite{SanzCillero:2005ef,Jamin:2008rm}.
In addition we also request that form factors given by those Green functions behave smoothly at high-$q^2$ transfer as partonic QCD demands \cite{Brodsky:1973kr,Lepage:1980fj}.
This procedure establishes a series of relations (not necessarily linear) between the couplings~:
\begin{equation} \label{eq:cons}
 f_j\left( \{ \alpha_i \} \right) \, = \, 0 \, ,
\end{equation}
that provide short-distance information on $\{ \alpha_i \}$. In the most probable situation
we have $j < i$ and then our knowledge on the couplings in ${\cal L}_I$ in Eq.~(\ref{eq:li}) is only
partial and we may complement the constraints by taking into account information from the phenomenology
of physical processes.

\subsection{R$\chi$T~: State of the art}

As commented above most of the applications from R$\chi$T are carried out with a Lagrangian involving only
the lightest nonet of scalar, pseudoscalar, vector and axial-vector resonances and corresponding external 
currents. Recently the extension to tensor resonances \cite{Ecker:2007us}
has also been carried out. To introduce heavier spectrum than the lightest nonets of resonances is a complicated
task that involves many more couplings. Hence little improvement in this aspect has been considered
\cite{Mateu:2007tr}.
\par
Most of the applications of R$\chi$T have taken place in the leading $N_C \rightarrow \infty$ limit, i.e. at tree
level. The next-to-leading order in the $1/N_C$ expansion involves complicated one-loop calculations in a 
non-renormalizable theory. However is still feasible to perform and is not only of interest for the 
phenomenology but also because it allows us a better understanding of the foundations of the theory.
A one-loop renormalization procedure including scalar and pseudoscalar resonances was
performed in Ref.~\cite{Rosell:2005ai}. The resonance saturation of the chiral LECs at one-loop level
has already been clarified \cite{Cata:2001nz,Portoles:2006nr} and fully explained \cite{Rosell:2009yb}.
Other related issues have also been addressed \cite{Rosell:2006dt,Pich:2008jm,SanzCillero:2009ap,Golterman:2006gv}.
The determination of the pion vector form factor within this frame has been undertaken \cite{Rosell:2004mn}
showing that the study of observables at this order is feasible. Finally a study on the renormalization group
equations in R$\chi$T has been carried out \cite{SanzCillero:2009pt} and other questions on renormalization
have also been considered \cite{Xiao:2007pu,Kampf:2009jh}.
\par
The main application of R$\chi$T to processes dominated by the lightest nonet of resonances is undoubtedly the
hadronization of QCD vector and axial-vector currents, i.e. hadron production in the $e^+ e^-$ cross-section and
hadron tau decays. A lot of work has been dedicated to this goal and with excellent results
\cite{GomezDumm:2000fz,Dumm:2009va,Guerrero:1997ku,GomezDumm:2003ku,Dumm:2009kj} and there is still a lot to do.
\par
Finally, the determination of the chiral LECs is also an all-important goal of R$\chi$T and considerable
effort has been invested with notable results \cite{Pich:2008xj}. A better knowledge of the 
theory would allow a thorough understanding and computation of many $\chi$PT observables.

\subsection{R$\chi$T~: Problems, solutions, ways out ...}
Let us comment on the procedure just devised above, its possible problems and treatments.
\par
The matching procedure between our R$\chi$T theory and the large-$q^2$ behaviour of Green functions is not
what one would do in an EFT. Here one should match the Lagrangian at an intermediate energy region $\mu = \Lambda$ where both, the lower and the upper theories, are valid. In the R$\chi$T case we do not have
a higher energy theory that we can match due to hadronization issues~: indeed we have a lower theory of mesons
and an upper theory in terms of quark and gluon fields. Hence our match happens at  $\Lambda \rightarrow \infty$,
i.e. with partonic QCD. We expect that the information that the Lagrangian gets from the matching is still
sounded.
In the usual applications carried out until now one only includes the lightest multiplet of resonances in the
R$\chi$T Lagrangian. Hence one could doubt that the Lagrangian may be able to  match the $\Lambda \rightarrow \infty$
energy region. However this is not intrinsic to the theory, one can include as much multiplets as one wishes,
though the number of couplings increases and it is much more complicated to fix them \cite{Mateu:2007tr}.
\par
There is a direct relation between the concept of {\em simplest} operators ${\cal O}_i$ 
and the matching procedure mentioned before. As there is no counting in the construction of the phenomenological
Lagrangian of R$\chi$T one could argue that the structure of the operators is not well defined when we 
speak about the minimal number of derivatives. This is indeed the case. However it is a fact that a high
number of derivatives tends to spoil the high-energy behaviour given by the OPE expansion and, accordingly,
the matching procedure. Moreover it can also happen that with the initial Lagrangian the matching of the 
Green functions is just not possible and then additional operators, not following that rule of {\em simplicity},
should be added. As a rule of thumb, admittedly arbitrary but very fruitful phenomenologically, one considers
the {\em simplest} operators that allow a matching of the Green functions.
\par
From a quantum theory of fields point of view, a hadron resonance field is not an asymptotic state as it decays
strongly. Detectors only track their decay products. This issue stays at the origin  of an ambiguity in the 
identification of the outgoing or ingoing resonance states with the fields in a Lagrangian. Then to use R$\chi$T
to study processes of production or decay of resonances
always could bring in that issue, notably if we are considering wide resonances. Accordingly we think that R$\chi$T is rigorously stated only when applying it to processes
whose dynamics is driven by resonances which are not outgoing or ingoing states.
\par
An additional issue is that of the off-shell widths. At leading order in the $1/N_C$ expansion
resonances have zero-width, as those are next-to-leading order effects. However when we apply, for instance,
R$\chi$T at tree level to the study of tau decays it is clear that resonances do resonate and, therefore, 
zero widths bring divergencies into the calculations. Hence we do have to include widths for the resonances. Moreover
if these are wide, like $\rho(770)$ or $\mathrm{a}_1(1260)$, a constant width is not a proper implementation and
we have to consider its off-shell behaviour. We have addressed this issue within the same R$\chi$T framework
\cite{GomezDumm:2000fz,Dumm:2009va}.
\par
Finally it has been observed that in some cases the constraints in Eq.~(\ref{eq:cons}) brought by
matching procedures to different Green functions or the use of different observables, for instance, can give conditions on the couplings that are at variance \cite{Bijnens:2003rc}. We still do not have a clear
understanding of the origins of that problem.

\section{Form factors in $\tau \rightarrow \pi \pi \pi \nu_{\tau}$ decays}

\begin{figure}[!t]
  \includegraphics[height=.25\textheight]{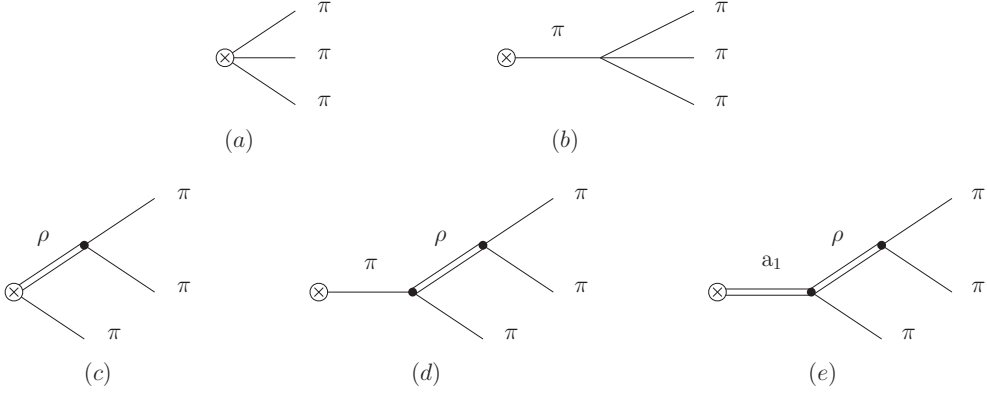}
  \caption{\label{fig:1} \small{Diagrams contributing to the axial-vector form factors $F_i$.}}
\end{figure}
As an example of application let us consider the decay of the tau lepton into three pions. Here we will
focus on the results of the approach and we refer to Refs.~\cite{Dumm:2009va,GomezDumm:2003ku} for details.
\par
In the Standard Model, the decay amplitudes for $\tau^- \rightarrow \pi^+
\pi^- \pi^- \nu_{\tau}$ and $\tau^- \rightarrow \pi^- \pi^0 \pi^0
\nu_{\tau}$ decays can be written as
\begin{equation}
{\cal M}_{\pm}  \, = \,  - \,
\frac{G_F}{\sqrt{2}} \, V_{ud} \, \bar u_{\nu_\tau}
\gamma^\mu\,(1-\gamma_5) u_\tau\, T_{\pm \mu} \; ,
\end{equation}
where $V_{ud}\simeq \cos\theta_C$ is an element of the
Cabibbo-Kobayashi-Maskawa matrix, and $T_{\pm\mu}$ is the hadron matrix
element of the axial-vector QCD current $A_{\mu}$,
\begin{equation} \label{eq:tmu1}
T_{\pm \mu}(p_1,p_2,p_3) \,  =  \,
 \langle  \pi_1(p_1)\pi_2(p_2)\pi^{\pm}(p_3)  |\, A_\mu\,
 e^{i {\cal L}_{QCD}} | 0  \rangle \, ,
\end{equation}
as there is no contribution of the vector current to these processes in
the isospin limit. Outgoing states $\pi_{1,2}$ correspond here to $\pi^-$
and $\pi^0$ for upper and lower signs in $T_{\pm\mu}$, respectively. The
hadron tensor can be written in terms of three form factors, $F_1$, $F_2$
and $F_P$, as~:
\begin{equation}
T^{\mu} \; = \; \left(  g^{\mu \nu} \, - \,
\frac{Q^{\mu} Q^{\nu}}{Q^2}  \right) \, (  p_1 - p_3  )_{\nu} \; F_1 \, + \,  
\left(  g^{\mu \nu} \, - \,
\frac{Q^{\mu} Q^{\nu}}{Q^2}  \right) \, (  p_2 - p_3  )_{\nu} \; F_2 \, + \,
Q^\mu\,F_P \; \; ,
\label{tmu}
\end{equation}
where $Q^\mu  =   p_1^\mu + p_2^\mu + p_3^\mu$.
In this way, the terms involving the form factors $F_1$ and $F_2$ have a
transverse structure in the total hadron momenta $Q_{\mu}$, and drive a
$J^P=1^+$ transition. Meanwhile $F_P$ accounts for a $J^P=0^-$ transition
that carries pseudoscalar degrees of freedom and vanishes with the square
of the pion mass. Its contribution to the spectral function of $\tppp$
goes like $m_{\pi}^4/Q^4$ and, accordingly, it is very much suppressed
with respect to those coming from $F_1$ and $F_2$. We do not consider it.
\par
The evaluation of the form factors $F_1$ and $F_2$ can be carried out within
R$\chi$T. The Lagrangian includes the original terms in R$\chi$T 
\cite{Ecker:1988te} together with those that satisfy the OPE expansion of
the $VAP$ Green function \cite{Cirigliano:2004ue}. Moreover we also include
the constraint of the smooth asymptotic behaviour of the $F_i$ form factors.
The different contributions \cite{Dumm:2009va,GomezDumm:2003ku} 
correspond to the diagrams in Fig.~\ref{fig:1}. Finally
the $\rho(770)$ and  $\mathrm{a}_1(1260)$ off-shell widths according
to R$\chi$T have also been implemented \cite{GomezDumm:2000fz,Dumm:2009va}.
\par
\begin{figure}[!t]
\includegraphics[scale=0.46,angle=-90]{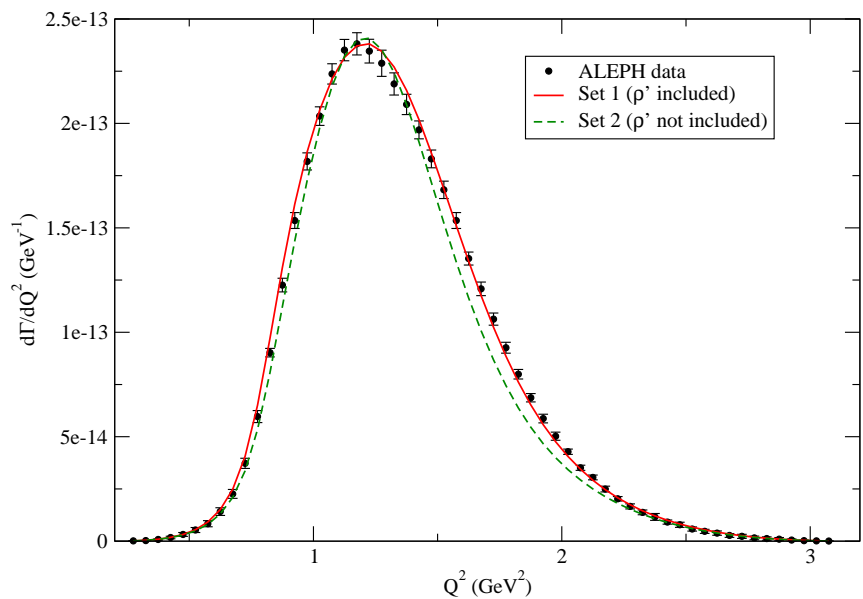}
\caption{\label{fig:2} \small{Comparison between the theoretical
$M_{3\pi}^2$-spectra of the $\tau^- \rightarrow \pi^+ \pi^- \pi^-
\nu_{\tau}$  with ALEPH data \cite{Barate:1998uf}. See details and explanations in Ref.~\cite{Dumm:2009va}}}
\end{figure}
We have considered that in $\tppp$ decays the $\rho(1450)$ state belonging to a heavier nonet of vector
resonances might play a role to explain the spectrum. However we have chosen not to include 
a new nonet in the R$\chi$T Lagrangian but as a modification of the $\rho$ Breit-Wigner weighted by
a new unknown parameter. In Fig.~\ref{fig:2} we show the comparison of ALEPH data \cite{Barate:1998uf} with the
R$\chi$T approach with and without the $\rho(1450)$ contribution. We see that the result is fairly good for
$M_{\mathrm{a}_1} \simeq 1.120 \, \mbox{GeV}$.

\section{Conclusions}

Resonance Chiral Theory is a systematic quantum field theory procedure that is able to handle those processes which dynamics is dominated
by the lightest unflavoured meson resonances. It has proven to be very fruitful in the analyses of
hadronization processes such as hadronic tau decays or the hadron $e^+ e^-$ 
cross-section for $\sqrt{s} \lsim 2 \, \mbox{GeV}$. It implements the constraints of chiral and 
unitary symmetry, the OPE leading behaviour of Green functions in partonic QCD and the smooth 
bearing of form factors at $q^2 \rightarrow \infty$. Finally it also represents a setting within
the $1/N_C$ expansion that rules its perturbative expansion. 
\par
R$\chi$T is also able to get information on the LECs in $\chi$PT. This is an essential
task because the predictability of the many ${\cal O}(p^6)$ $\chi$PT calculations relies on our knowledge
of the LECs that show up in the observables.


\begin{theacknowledgments}
I wish to thank the organizers of Chiral10 for their kind invitation to give
this talk. I also would like to thank I.~Rosell, P.~D.~Ruiz Femen\'{\i}a and J.~J.~Sanz Cillero for their comments
on the draft of this note. This work has been supported in part by the EU MRTN-CT-2006-035482 (FLAVIAnet), 
by MEC (Spain) under Grant No. FPA2007-60323, by the Spanish Consolider-Ingenio
2010 Programme CPAN (CSD2007-00042), and by Generalitat Valenciana under Grant No.
PROMETEO/2008/069.
\end{theacknowledgments}



\bibliographystyle{aipproc}   

\bibliography{rcht}

\IfFileExists{\jobname.bbl}{}
 {\typeout{}
  \typeout{******************************************}
  \typeout{** Please run "bibtex \jobname" to optain}
  \typeout{** the bibliography and then re-run LaTeX}
  \typeout{** twice to fix the references!}
  \typeout{******************************************}
  \typeout{}
 }

\end{document}